\documentclass[jcp,twocolumn,
superscriptaddress,showpacs,floatfix]{revtex4-1}

\usepackage{bm}
\usepackage{amsmath}
\usepackage{amssymb}
\usepackage{graphicx}
\usepackage{dcolumn}
\usepackage{color}
\usepackage{MnSymbol}
\usepackage{soul,xcolor}
\setstcolor{red}

\usepackage{xcolor}\definecolor{ao}{rgb}{0.0,0.0,1.0}

\definecolor{br}{rgb}{1.0, 0.22, 0.0}

\usepackage{epsfig}

\def\m{{\mbox{\boldmath{$\mu$}}}}

\input{epsf}
\begin{document}
\title{
ac transport and full-counting statistics of molecular junctions\\
in the weak electron-vibration coupling regime}

\author{A. Ueda}
\affiliation{Faculty of Pure and Applied Sciences, Division of Applied Physics, University of Tsukuba,
 Tsukuba, Ibaraki, 305-8573, Japan}
 
\author{Y. Utsumi}
 \affiliation{Department of Physics Engineering, Faculty of Engineering, Mie University, Tsu, Mie, 514-8507, Japan }

\author{Y. Tokura}
\affiliation{Faculty of Pure and Applied Sciences, Division of Physics, University of Tsukuba, Tsukuba, 305-8573, Japan}

\author{O. Entin-Wohlman}
\affiliation{Physics Department, Ben Gurion University, Beer Sheva 84105, Israel}
\affiliation{Raymond and Beverly Sackler School of Physics and Astronomy, Tel Aviv University, Tel Aviv 69978, Israel}
 \email{orawohlman@gmail.com}

\author{A. Aharony}
\affiliation{Physics Department, Ben Gurion University, Beer Sheva 84105, Israel}
 \affiliation{Raymond and Beverly Sackler School of Physics and Astronomy, Tel Aviv University, Tel Aviv 69978, Israel}

\date{\today}

\begin{abstract}

The coupling of the charge carriers passing through 
a molecule bridging two bulky conductors with  local vibrational modes of the molecule, gives rise to  distinct features in  the electronic transport properties on one hand, and to nonequilibrium features in   the vibrations' properties, e.g., their population,   on the other. Here we explore theoretically  a generic  model for  a molecular junction biased by an arbitrary dc  voltage in the weak-coupling regime. We analyze the 
signature of  the   electron-vibration interaction on 
the full-counting statistics of the current fluctuations  (i.e., the cumulant generating-function of the current correlations); 
we give a detailed account of the response to an ac field exerted on the junction (on top of the dc bias voltage); we study the nonequilibrium distribution of the vibrational modes and the  fluctuations they cause in the displacement of the molecule center of mass.
The calculations use  the technique of nonequilibrium Green's functions, and treat   the electron-vibration coupling in perturbation theory,  within the random-phase approximation when required. 

\end{abstract}


\maketitle

\section{Introduction}
\label{Intro}

Molecular junctions, which are metallic electrodes bridged by a single    molecule   (or a few molecules), are currently a subject of considerable interest due to their possible applications in molecular electronics. \cite{Cuevas}  The particular feature of these  setups, that distinguishes them from e.g.,  quantum-dot or quantum-wire junctions, is 
the coupling between the motion  of  the molecule's vibrations, e.g., those of  the center-of-mass,  and the single-electron tunneling.  Nano  electro-mechanical vibrations were indeed detected   in a single-C$_{60}$ transistor. \cite{Park} 
Early experiments, achieving   almost a perfect transmission via a single molecule,  were carried out on 
break-junction devices bridged by H$_{2}$. \cite{Smit}
Conductances comparable to those of metallic atomic junctions were detected also for benzene molecules coupled to platinum leads. \cite{Kiguchi}
These are just a few examples of the huge body of experimental results concerning  electric transport through molecular junctions. However, these devices have other attributes.
Molecular
junctions are particularly useful for studying electro-mechanical interactions in the quantum regime. When
the bias voltage across a molecular junction exceeds the
energy of a given mode of vibration, then  that mode can be
excited, at low temperatures, by the electrons injected from the source electrode.  This  results in an additional 
contribution to the electric current.
Whether this inelastic event  increases or decreases  the measured differential conductance is an intriguing question.
At sufficiently strong electron-vibration coupling the current flow at low biases is found to be suppressed, a phenomenon termed the  ``Franck-Condon blockade".
\cite{Leturcq2009}
On the other hand, 
a clear crossover  between enhancement and reduction of the dc conductance    was detected in shot-noise measurements, in a H$_{2}$O molecular junction,   \cite{Tal2008} and in gold nanowires. \cite{Agrait}
Beside electronic transport, other properties of molecular junctions are being explored.   \cite{Aradhya} 
Inelastic neutron tunneling spectroscopy \cite{Scheer} and Raman response \cite{Natelson} were used to study the molecular conformation, and other characteristics of the junction itself. The electron-vibration coupling also induces renormalization, damping,   \cite{Koch.2006} and heating of the vibrational modes, 
\cite{Kaasbjerg}        whose study could explain certain features in the Raman spectroscopy of 
OPV3 junctions. \cite{Ward}
Interestingly
enough, the ability to measure thermoelectric effects in molecular junctions
provides a tool to determine   the electronic structure of the molecule, for instance, by monitoring the Seebeck coefficient (for a recent review, see Ref. \onlinecite{Rincon}).
Transport through molecular bridges coupled to metallic electrodes has been also exploited to investigate electronic correlations, e.g., the Kondo effect. \cite{Rakhmilevitch}

The theoretical analysis of  transport through molecular junctions has been carried out by  a vast variety of methods. These include {\it ab initio} 
computations (see, e.g., Ref. \onlinecite{Paulsson,French,Liu,Bai}), 
mixed  quasi-classical and semiclassical approaches (see, e.g., Ref. \onlinecite{Lu,Rabani}), calculations based on scattering theory, \cite{Caspary,Jom,Zimbovskaya}  constructions of quantum master equations,  \cite{Harbola} using  real-time path-integrals combined with  Monte Carlo computations, \cite{Muhlbacher} and more.

In this paper we consider  the effect of the electron-vibration coupling on transport properties  at an arbitrary  bias voltage, i.e., when    transport is   beyond the linear-response regime. The coupling of the vibrations with the charge carriers  naturally involves also inelastic processes.  At very low temperatures, as considered in this paper, real  inelastic scattering events are feasible when the bias voltage exceeds the threshold of the vibrational modes' energy. One therefore expects unique features at bias voltages around this energy. The application of an additional ac field as  considered below gives rise to an interplay between the ac frequency and the frequencies  of the vibrational modes. The focus of our paper is the study of the dynamics of the charge carriers and the oscillations of the center of mass of the molecule over a wide range  of bias voltages, vibrational frequencies, and ac frequencies.

Under these circumstances, a suitable method to use is that of the nonequilibrium  Green's functions, i.e., the Keldysh technique. \cite{Keldysh}    We apply this technique  to  the 
ubiquitous simple model for molecular junctions, which
replaces the molecule by a quantum dot with a single localized level attached to two electronic reservoirs. 
Electrons residing on the level exchange energy with Einstein vibrations (or optical phonons), of frequency $\omega_{0}$, resulting from oscillations of the junction, as represented by the localized level.
Even for  a weak electron-phonon coupling, this model, which has been pursued  for more than a decade,  
produces intriguing features in the  transport properties. 
\cite{Mitra,Egger,OEW.2009} As in our previous works on this topic,  \cite{OEW.2009,OEW.2010t, AU.2011,OEW.2012,AU.2013,Imamura}
we confine ourselves to this regime, treating the electron-vibration interaction in the lowest possible order in the coupling energy.  \cite{Egger,OEW.2009,Haupt,Avriller,Schmidt,Urban} Note, however, that this procedure is rather delicate and care must be taken in exploiting it (see, e.g.,  the discussion in Sec. \ref{fcs}).
The limit where the vibrations are strongly coupled to the charge carriers, \cite{Galperin,Souto,Simine,Maier,Schaller} and the effect of electron-electron correlations,  \cite{Martin_Rodero,Cornaglia,Erpenbeck,Chen}  are beyond the scope of this paper.

While considerable theoretical effort has been devoted to the study of dc transport, less attention has been paid to the response of molecular junctions to  a frequency-dependent  electric field. The ac conductance for tunneling through  an arbitrary interacting quantum dot was analyzed in Ref. \onlinecite{Kubala}, and polaronic  effects were considered in Ref. \onlinecite{Ding}, assuming that the vibrational modes are equilibrated on a time scale shorter than the transit time of the electrons through the junction. 
We focus on the situation where the vibrational modes are equilibrated via their interaction with the charge carriers; in particular we explore the  effect on the full counting statistics,  and the modifications introduced by an ac field in the nonequilibrium distribution of the vibrational modes, together with  its effect on the oscillations of the center of mass of the molecule.

Our paper is organized as follows. In Sec. \ref{model} we describe the model used for the calculations. To set the stage for the discussion of the ac current in the presence of 
 arbitrary dc voltages,   in linear response to an ac field, we review in  Sec. \ref{dc} certain properties of the dc current at finite voltages. Section \ref{ac} contains a detailed analysis of the ac response of the junction;  particular attention is paid to the dependence of the response coefficient on the ac frequency. This part of the calculation involves the consideration of various diagrams, whose individual contributions to the ac  transport coefficient is not easy to anticipate. We list in Appendix \ref{detC}  these diagrams, their detailed expressions,  and display plots of their separate contributions to the ac response. The investigation of the effect of an ac field on the dynamics on the junction is continued in Sec.  \ref{vib},  where we study  two correlation functions of the vibrations, the first is related to the nonequilibrium distribution of the vibrations, and the other to the fluctuation in the displacement  of the harmonic oscillator representing the  junction. The dependence of the two quantities on the ac frequency is analyzed. In particular we find that the phase delay of the fluctuation shows a structure at two specific values of that frequency, one which can be explained  by considering a classical driven oscillator, and another which cannot; it stems from  two-vibration scattering by the charge carriers, and thus exemplifies a quantum electron-mechanical effect.
Section \ref{fcs} reviews our recent results for the cumulant generating-function of our model and dwells in particular on its modifications due to the electron-vibration coupling and the nonequilibrium distribution of the vibrations.    An intriguing relation between the full counting statistics and the theory of thermodynamic phase transitions is pointed out. Section \ref{conclu} summarizes briefly our work.

\begin{figure}[htp]
\includegraphics[width=6cm]{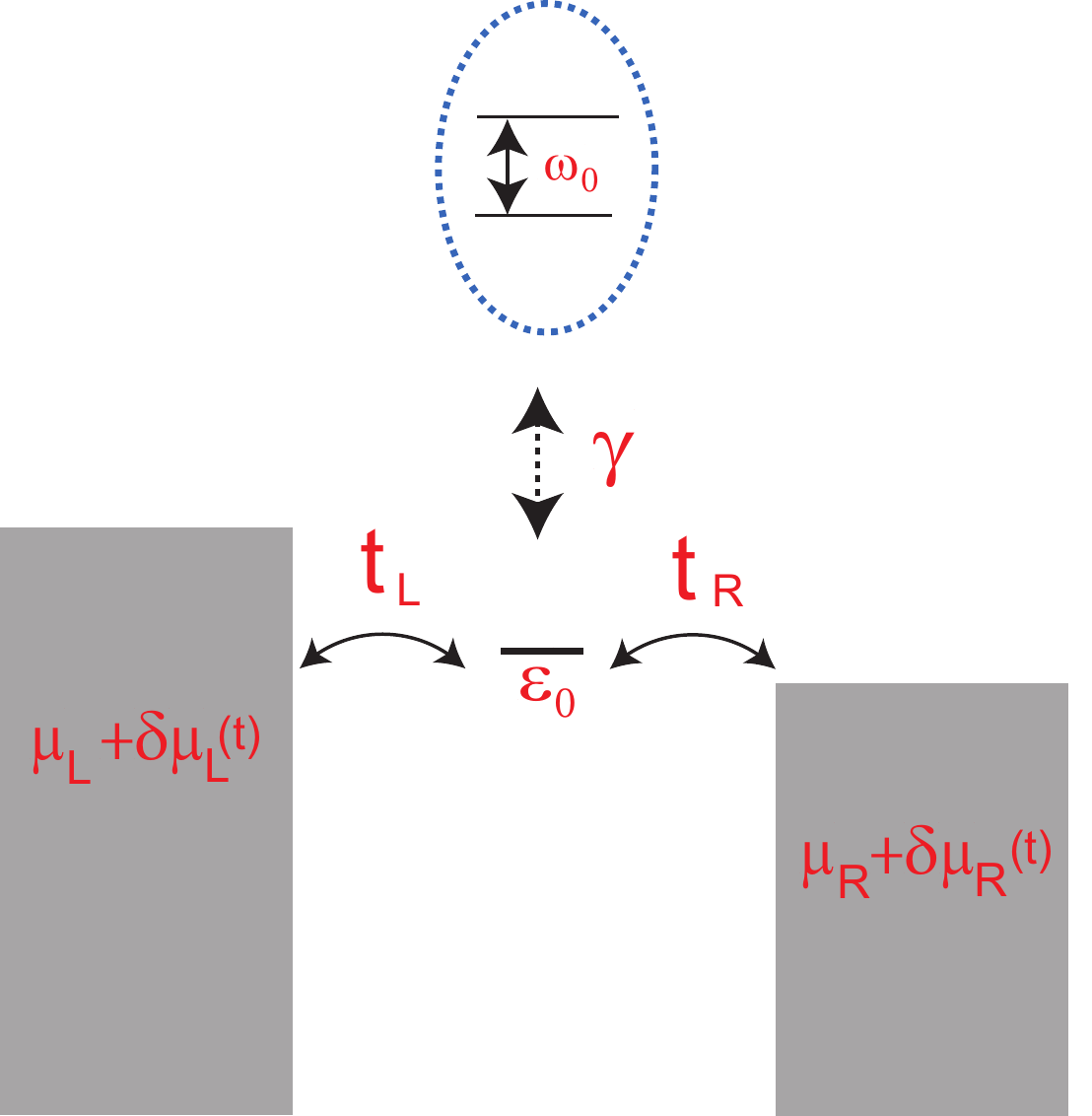}
\caption{(color online) Illustration of the model used in the calculation. An electronic localized level, of energy $\epsilon_{0}$, is coupled to two bulky electronic reservoirs, held at two different chemical potentials, $\mu_{L(R)}+\delta\mu_{L(R)}(t)$ for the left (right) electrode.    The charge carriers exchange energy with Einstein vibrations, of frequency $\omega_{0}$; the coupling energy of the electrons with the vibrations is denoted $\gamma$. The difference $\mu_{L}-\mu_{R}=eV$  is the bias voltage multiplied by the unit of charge; $\delta\mu_{L(R)}(t)$  are monochromatic  ac fields of frequency $\omega_{\rm ac}$ applied to the junction.  These are treated in the linear-response approximation. }
\label{Fig1}
\end{figure}

\section{Electric current and vibrational modes dynamics}
\label{basic}

\subsection{The model Hamiltonian and the electric current}
\label{model}
 
The  model we use is depicted in Fig. \ref{Fig1}: a localized electronic level,  of energy $\epsilon_{0}$,  is coupled to two electronic electrodes, which are held at two different chemical potentials, $\mu_{L}+\delta\mu_{L}(t)$ and $\mu_{R}+\delta\mu_{R}(t)$.
An ac field of frequency $\omega_{\rm ac}$  applied to the junction is represented by a periodic time-dependence of the chemical potentials, $\delta\mu_{L(R)}(t)$. 
Specifically we choose 
\begin{align}
\mu^{}_{L}+\delta\mu^{}_{L}(t)&=\mu+eV/2+\delta\mu^{}_{L}\cos (\omega^{}_{\rm ac}t)\ ,\nonumber\\
\mu^{}_{R}+\delta\mu_{R}^{}(t)&=\mu-eV/2+\delta\mu^{}_{R}\cos (\omega^{}_{\rm ac}t)\ ,
\label{mu}
\end{align}
where $V$ is the bias voltage, and $\mu$ is the  common chemical potential of the electrodes.
An electron on the level is coupled to  local Einstein vibrations; this coupling  induces fluctuations in  the level energy. \cite{Mitra,Holstein,Egger,OEW.2009,Galperin.2007}
The model Hamiltonian is
\begin{align}
{\cal H}={\cal H}^{}_{\rm lead}+{\cal H}^{}_{\rm mol}+{\cal H}^{}_{\rm ph}+{\cal H}^{}_{\rm tun}\ .
\label{HAM}
\end{align}
The two electronic electrodes (assumed to be identical except being kept at different chemical potentials) are represented by free-electron  gases, 
\begin{align}
{\cal H}^{}_{\rm lead}&=\sum_{k}(\epsilon^{}_{k}-\mu-eV/2)c^{\dagger}_{k}c^{}_{k}\nonumber\\
&+
\sum_{p}(\epsilon^{}_{p}-\mu+eV/2)c^{\dagger}_{p}c^{}_{p}\ ,
\label{hamlead}
\end{align}
where $c^{\dagger}_{k(p)}$ and $c^{}_{k(p)}$
denote the creation and annihilation operators 
of an electron of momentum $k(p)$ and energy  $\epsilon_{k(p)}$ in the left (right) electrode, respectively.
The Hamiltonian of the localized level reads
\begin{align}
{\cal H}^{}_{\rm mol}=[\epsilon^{}_{0}+\gamma (b+b^{\dagger})]
c^{\dagger}_{0}c^{}_{0}\ ,
\label{hammol}
\end{align}
with the creation and annihilation operators $c^{\dagger}_{0}$ and $c^{}_{0}$, respectively, for an electron on the localized level. The second term in the square brackets is the (linear) electron-vibration coupling: the creation and annihilation operators of the Einstein   vibrations are $b^{\dagger}_{}$ and $b$, respectively, and the electron-vibration coupling energy is $\gamma$. The vibrational modes obey the Hamiltonian
 \begin{align}
 {\cal H}^{}_{\rm ph}=\omega^{}_{0}b^{\dagger}b\ .
 \label{hamph}
 \end{align} 
 (We use units in which $\hbar=1$.)
Our calculations are carried out in second-order perturbation theory in the electron-vibration coupling, i.e., we keep terms up to order $\gamma^{2}$. However, in the absence of the ac field, this approximation is not sufficient for the determination of the vibrations' population [see the discussions following Eq. (\ref{pl}) and in Sec. \ref{fcs}].

The tunneling Hamiltonian connecting the localized level with the left (right) electrode is specified by the tunneling amplitude $t_{L(R)}$,  and is written in  the form
 \begin{align}
 {\cal H}^{}_{{\rm tun},\pm}
 =\sum_{k}t^{}_{L}e^{\pm i\lambda^{}_{L}}c^{\dagger}_{k}c^{}_{0}+
 \sum_{p}t^{}_{R}e^{\pm i\lambda^{}_{R}}c^{\dagger}_{p}c^{}_{0} +{\rm H.c.}\ . 
 \label{hamtun}
 \end{align}
Here, $\lambda_{L(R)}$ are the counting fields. \cite{Utsumi.2006,AU.2013}  These are introduced to facilitate the calculation of the full-counting statistics (Sec. \ref{fcs}). In the long-time limit, the cumulant generating-function
 depends only on the difference of the two, i.e., 
 on
 \begin{align}
 \lambda=\lambda^{}_{L}-\lambda^{}_{R}\ .
 \end{align}
 As  $
\lambda^{}_{L}+\lambda^{}_{R}$ 
 counts the number of electrons flowing into the localized level, the fact that the cumulant generating-function depends solely on $\lambda$ implies charge conservation. For the calculation of the  response of the system to the chemical potentials  we set $\lambda_{L}=\lambda_{R}=0$.

\begin{widetext}
The average (time-dependent) current emerging from the left electrode is 
\begin{align}
I^{}_{L}(t) &=-e\Big \langle \frac{d}{dt}\sum_{k}c^{\dagger}_{k}c^{}_{k}\Big \rangle\nonumber\\
&=e{\rm Re}\Big (\int dt'\sum_{k}|t^{}_{L}|^{2}\Big [G^{r}_{00}(t,t')g^{<}_{k}(t',t)+
G^{<}_{00}(t,t')g^{a}_{k}(t'-t)-
g^{r}_{k}(t-t')G^{<}_{00}(t',t)
-g^{<}_{k}(t,t')G^{a}_{00}(t',t)\Big ]\Big )\ .
\label{IL}
\end{align}
 \end{widetext}
The current emerging from the right electrode is obtained from Eq. (\ref{IL})
by the replacements $L\rightarrow R$ and $k\rightarrow p$. 
Here, 
$G_{00}$ is the Green's function on the localized level in the presence of the coupling to the electrodes, the ac  fields $\delta\m_{L(R)}$,  and the electron-vibration interaction. The retarded and the advanced electron Green's functions are  
\begin{align}
G^{r/a}_{00}(t,t')=\mp i\Theta (\pm t\mp t')\langle\{c^{}_{0}(t),c^{\dagger}_{0}(t')\}\rangle\ ,
\label{Gra}
\end{align}
and the Keldysh lesser Green's function \cite{Keldysh} is
\begin{align}
G^{<}_{00}(t,t')=i\langle c^{\dagger}_{0}(t')c^{}_{0}(t)\rangle \ .
\end{align} 
The Green's functions  on the decoupled electrodes are denoted by the lowercase $g_{k(p)}$, with
 \begin{align}
g^{r/a}_{k(p)}(t,t')=\mp i\Theta (\pm t\mp t')\langle\{c^{}_{k(p)}(t),c^{\dagger}_{k(p)}(t')\}\rangle\ ,
\end{align}
and 
\begin{align}
g^{<}_{k(p)}(t,t')=i\langle c^{\dagger}_{k(p)}(t')c^{}_{k(p)}(t)\rangle \ .
\end{align}

We consider the response of the junction to the ac fields in {\em linear response}. That is, while the bias voltage is arbitrary, $\delta\mu_{L(R)}$ [Eq. (\ref{mu})] are assumed to be small. To this end, we use the expansion
\begin{align}
\sum_{k(p)}g^{<}_{k(p)}(t,t')&=i\nu\int
d\epsilon^{}_{k(p)} f^{}_{L(R)}(\epsilon^{}_{k(p)})e^{-i\epsilon^{}_{k(p)}(t-t')}\nonumber\\
&\times\exp\Big [i\int_{t'}^{t}dt^{}_{1}\delta\mu^{}_{L(R)}e^{i\omega^{}_{\rm ac}t^{}_{1}}\Big ]\ ,
\label{gele}
 \end{align}
where 
\begin{align}
f^{}_{L(R)}(\omega )=[e^{\beta(\omega -\mu\mp eV/2)}+1]^{-1}_{}
\end{align} 
is the Fermi distribution function
 of the electrons in the left (right) electrode, and $\beta=1/(k_{\rm B}T)$ is the inverse temperature. In Eq. (\ref{gele}),  $\nu$ is the density of states of the electron gases in the electrodes, at the common chemical potential.   [Note that $\delta\mu_{L(R)}\cos (\omega_{\rm ac}t) $ is replaced in the calculations  by 
$\delta\mu_{L(R)}\exp (i\omega_{\rm ac}t) $. \cite{AU.2011}]
Using the expansion (\ref{gele}), 
\begin{align}
I^{}_{L}(t)\approx I^{}_{L}+C^{}_{LL}(\omega^{}_{\rm ac})e^{i\omega^{}_{\rm ac}t}\delta\mu^{}_{L}+
C^{}_{LR}(\omega^{}_{\rm ac})e^{i\omega^{}_{\rm ac}t}\delta\mu^{}_{R}\ ,
 \label{ilac}
\end{align}
 where $I^{}_{L}$ is the dc current flowing from the left electrode.  
The other  terms on the right-hand side of Eq.  (\ref{ilac}) describe the ac current emerging from the left electrode in linear response (with respect to $\delta\mu_{L}$ and $\delta\mu_{R}$).

We adopt the definition 
\begin{align}
I(t)=[I^{}_{L}(t)-I^{}_{R}(t)]/2\ ,
\end{align}
for the current flowing through the junction. 
In Sec. \ref{dc} we examine the dc current; Sec. \ref{ac} analyzes the response to the ac field,
assuming for concreteness that $\delta\mu_{L}+\delta\mu_{R}=0$, and confining the discussion to a spatially symmetric junction. With these assumptions, the response to the ac field is given by a single coefficient  
\begin{align}
C(\omega^{}_{\rm ac})=[C^{}_{L}(\omega^{}_{\rm ac})+C^{}_{R}(\omega^{}_{\rm ac})]/2\ .
\label{Ct}
\end{align}
(Since part of the coefficient $C_{LL(RR)}$ is cancelled against $C_{LR(RL)}$, the notations are modified. \cite{AU.2011})  
We remind the reader that while the ac field is treated in linear response, the bias voltage $V$ is arbitrary.


 \subsection{The dc current}
 \label{dc}

In the absence of the ac field (but in the presence of a constant bias voltage), the dc current through the junction can be written in the form  (see, for instance,  Ref. \onlinecite{OEW.2009})
\begin{align}
I&=(I^{}_{L}-I^{}_{R})/2\nonumber\\
&=e\frac{\Gamma^{}_{L}\Gamma^{}_{R}}{\Gamma}\int\frac{d\omega}{2\pi} {\rm Im}G^{a}_{00}(\omega)[f^{}_{L}(\omega )-f^{}_{R}(\omega )]\ .
\label{idc}
\end{align} 
Due to the coupling with the leads, the localized level becomes a resonance, whose width is $\Gamma$, 
\begin{align}
\Gamma=\Gamma^{}_{L}+\Gamma^{}_{R}\ ,\ \ \ \Gamma^{}_{L(R)}=2\pi\nu|t^{}_{L(R)}|^{2}\ ,
\label{Gamma}
\end{align}
 where $\Gamma_{L}$  ($\Gamma_{R}$) is the partial width arising from the coupling to left (right) electrode. These partial widths determine the transmission ${\cal T}$ of the junction at the common chemical potential of the leads, 
 \begin{align}
{\cal T}=\frac{\Gamma^{}_{L}\Gamma^{}_{R}}{(\mu-\epsilon^{}_{0})^{2}+(\Gamma/2)^{2}} \ .
\label{tr}
\end{align}
The Green's function  $G_{00}^{a}(\omega)$ is the Fourier transform of $G_{00}^{a}(t-t')$, Eq. (\ref{Gra}) [note that at steady state the functions in Eq. (\ref{Gra}) depend only on the time difference];  
to second order in the electron-vibration coupling $\gamma$,  it reads \cite{OEW.2009}
 \begin{align}
 G^{r/a}_{00}(\omega)=\Big [\omega -\epsilon^{}_{0}\pm i\Gamma/2-\Sigma^{r/a}_{\rm Har}-\Sigma^{r/a}_{\rm Ex}(\omega )\Big ]^{-1}\ . 
 \label{gra}
 \end{align}
 Here $\Sigma_{\rm Har}$ and $\Sigma_{\rm Ex}(\omega)$
constitute the self-energy due to that interaction. The  Hartree term of the interaction gives rise to a frequency-independent self energy, 
 \begin{align}
 \Sigma^{r/a}_{\rm Har}=-i\gamma^{2}\int\frac{d\omega}{2\pi} 
G^{(0)<}_{00}(\omega ) D^{r/a}_{}(0)\ ,
\label{har}
\end{align}
and the exchange term is
\begin{align}
\Sigma^{r/a}_{\rm Ex}(\omega)&=i\gamma^{2}\int\frac{d\omega'}{2\pi}\Big (
G^{(0)<}_{00}(\omega -\omega ')D^{r/a}_{}(\omega ')\nonumber\\
&+G^{(0)r/a}_{00}(\omega -\omega ')[D^{<}_{}(\omega ')\pm D^{r/a}_{}(\omega ')]
\Big )\ .
\label{ex}
\end{align}
The electron Green's functions appearing in Eqs. (\ref{har}) and (\ref{ex}) are  zeroth-order in the electron-vibration coupling, as these self-energies are proportional to $\gamma^{2}$. The expressions for these Green's functions  are
\begin{align}
 G^{(0)r/a}_{00}(\omega)=\Big [\omega -\epsilon^{}_{0}\pm i\Gamma/2\Big ]^{-1}\ . 
 \label{G0ra}
 \end{align}
 and
\begin{align}
G^{(0)<}_{00}(\omega)&
=i[\Gamma^{}_{L}f^{}_{L}(\omega)+\Gamma^{}_{R}f^{}_{R}(\omega )]
G^{(0)a}_{00}(\omega)G^{(0)r}_{00}(\omega )
\ .
\label{G0l}
\end{align}
[The greater Green's function is obtained upon replacing $f_{L(R)}$ by $f_{L(R)}-1$.]
The Hartree term in the self energy is ignored hereafter, as it just produces a shift in the localized-level energy.

The expressions for the self-energies include also the vibration Green functions, $D$,  
\begin{align}
D^{r/a}_{}(t,t')&=\mp i\Theta(\pm t\mp t')\langle [b(t)+b^{\dagger}_{}(t),b(t')+b^{\dagger}(t')]\rangle\ ,\nonumber\\
D^{<}_{}(t,t')&=-i\langle 
[b(t')+b^{\dagger}_{}(t')][b(t)+b^{\dagger}_{}(t)]\rangle \  .
\label{Dcon}
\end{align}
At steady state, the Fourier transforms of the vibration Green's functions
are 
\begin{align}
D^{r/a}_{}(\omega)=\frac{2\omega^{}_{0}}{(\omega\pm i\eta)^{2}-\omega^{2}_{0}+2\omega^{}_{0}\Pi^{r/a}_{}(\omega)}\ ,
\label{dra}
\end{align}
for the retarded and advanced functions, and
\begin{align}
D^{<}_{}(\omega)=D^{r}_{}(\omega)\Pi^{<}_{}(\omega)D^{a}_{}(\omega)\ , 
\label{dl}
\end{align}
for the lesser Green's function.
In Eqs. (\ref{dra}) and (\ref{dl}),  $\eta$ characterizes the approach of the vibrations' distribution to its equilibrium value (see below), 
 $\Pi$ is the electron-hole propagator, 
\begin{align}
\Pi^{r/a}_{}(\omega)&=-i\gamma^{2}\int\frac{d\omega'}{2\pi}[G_{00}^{r/a}(\omega')G^{<}_{00}(\omega'-\omega )\nonumber\\
&+G^{<}_{00}
(\omega ')G^{a/r}_{00}(\omega' -\omega )]\ ,
\label{pra}
\end{align}
and
\begin{align}
\Pi^{<}_{}(\omega )=-i\gamma^{2}\int\frac{d\omega'}{2\pi}G^{<}_{00}(\omega ')G^{>}_{00}(\omega'-\omega )\ ,
\label{pl}
\end{align}
where $G^{>}_{00}$ is the greater Keldysh function. \cite{Keldysh}
Since the electron-hole propagator is proportional to $\gamma^{2}$, the Green's functions $G_{00}$  in Eqs. (\ref{pra}) and (\ref{pl}) are those of the interaction-free junction, Eqs. ({\ref{G0ra}) and (\ref{G0l}).

There is a subtle feature in this model, which distinguishes between the equilibration processes of the electrons and of the vibrational modes. While the electrodes serve as thermal baths for the electrons, where they acquire the Fermi distribution (different for the two electrodes), the model, as described in Sec. \ref{model},  does not specify the relaxation of the vibrational modes. 
In other words, whether or not the vibrations can equilibrate in a time much shorter than the transit time of an electron through the junction affects significantly the dynamics. 
When it is  plausible to assume that the distribution 
of the vibrational modes is determined  by their coupling to a phonon bath (e.g., the substrate on which the junction is lying), and is given by the Bose-Einstein distribution, then $\eta\gg {\rm Im}[\Pi^{a}(\omega)]$. In contrast, in the other extreme case,  ${\rm Im}[\Pi]$
dominates, and the vibrational modes' distribution is governed by the coupling to the charge carriers. 
\cite{OEW.2010} 

\begin{figure}[htp]
\includegraphics[width=8cm]{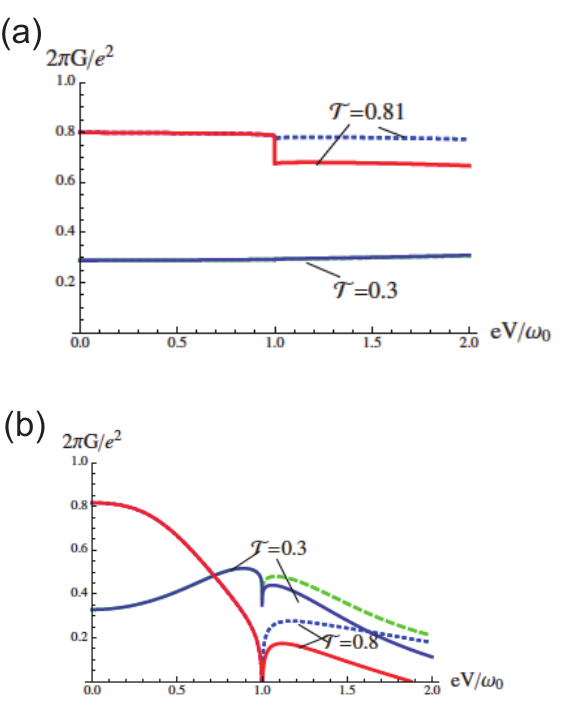}
\caption{(color online) The zero-temperature differential conductance $G$ (scaled by the quantum unit of the conductance) as a function of the bias voltage (scaled by $\omega_{0}/e$), for two values of the junction's transmission ${\cal T}$ (marked on the figure). The full lines [for ${\cal T}=0.81$ (red) and ${\cal T}=0.3$ (dark blue)] are the conductance 
when the vibrations' population is determined by the coupling to the charge carriers, and the dotted curves [for ${\cal T}=0.81$ (blue) and ${\cal T}=0.3$ (green)]   are for a Bose-Einstein population. The coupling energy $\gamma$ is $\omega_{0}/2$ and $\Gamma =6 \omega_{0}$ (a) and 
$\omega_{0}/6$ (b).
Adapted with permission from Phys. Rev. B {\bf 81}, 113408 (2010) (see Ref. \onlinecite{OEW.2010}); copyrighted by the American Physical Society.
}
\label{Fig2}
\end{figure}

Figure \ref{Fig2} displays
 the dependence of
the differential conductance, $G$,  on the bias voltage and on the
other parameters of the junction, and illustrates the difference between the two possibilities discussed above, i.e.,  $\eta\gg {\rm Im}[\Pi^{a}(\omega)]$ and
$\eta\ll {\rm Im}[\Pi^{a}(\omega)]$.
Figure \ref{Fig2}(a) shows the conductance of a junction tightly bound to the leads (the resonance width $\Gamma$   is relatively  large)
and hence the dwell time of the electrons is rather short.
Then, at relatively low values of the bare transmission, ${\cal T}=0.3$, 
there is no discernible modification in
the conductance which is about the same  for nonequilibrium vibrational modes' population and  for the equilibrium one. Figure \ref{Fig2}(b) depicts the situation when the resonance width is relatively small. There,  for higher values of the bare transmission, ${\cal T}=0.81$, 
the
step-down feature of the conductance at threshold for inelastic tunneling, $eV=\omega_{0}$,  is enhanced for nonequilibrium vibrations as
compared to the equilibrated ones.
Hence, 
when the bare transmission of the junction is high then the
conductance in the presence of nonequilibrium population is
lower than the one pertaining to the case of equilibrated vibrations. For low bare transmissions the difference is rather
small. One notes that the logarithmic singularity associated
with the real
part of the self-energy at the channel opening, \cite{OEW.2009}
is  manifested also when
the vibrations  are out of equilibrium, for high enough values of the bare transmission of the junction.
The fact that the changes in the differential conductance
are more pronounced for the larger 
values of $\omega_{0}/\Gamma$ is connected
 with the actual value of
the population.
The
electrons pump more and more excitations into the higher
vibrational states as the bias voltage increases and this pumping is more effective as the dwell time exceeds considerably
the response time of the oscillator.

\subsection{The ac current}

\label{ac}

In our previous work \cite{AU.2011}
we have found that the response of the charge carriers to a frequency-dependent  field, when the bias voltage vanishes,  can be enhanced (suppressed) by the coupling to the vibrations when the localized level lies below (above) the common chemical potential of the electrodes;
this behavior was attributed to the Franck-Condon blockade due to the Hartree term, Eq. (\ref{har}).
It was also found that the vertex corrections of that interaction induce an additional peak structure in  the response, 
which could be modified by tuning the width of the resonance, i.e., by varying $\Gamma_{L(R)}$.
This additional peak disappeared in a spatially-symmetric junction, for which $\Gamma_{L}=\Gamma_{R}$. 
In a more recent work,  \cite{Imamura}
we have included also the effect of a finite  bias voltage, assuming that the vibrations are equilibrated by their coupling to both a phonon bath [as described by the $\eta$, see the discussion following Eq. (\ref{pl})] and  the charge carriers 
(a possibility that exists at very low temperatures when $V\neq 0$). That analysis pertained  to a perfect junction for which ${\cal T}=1$ [see Eq. (\ref{tr})]. 
It was found that the response  can be enhanced by the coupling to the vibrations for sufficiently large values of the bias voltage. In addition, $C(\omega_{\rm ac})$ developed a sharp feature at $\omega_{\rm ac}=2\omega_{0}$; we discuss this structure below.  Here we extend that study for non perfect junctions, i.e., for ${\cal T}<1$.

The detailed  expressions for the Fourier transform of the transport coefficient $C(\omega_{\rm ac})$, are given in Appendix \ref{detC}, see also Ref. \onlinecite{Imamura}.
They are based on the expansion of the electron Green's function, in the absence of the coupling with vibrations, to linear order in the ac amplitude $\delta\mu=\delta\mu_{L}=-\delta\mu_{R}$, 
\begin{align}
G^{({\rm nint)}<}_{00}(t,t')=G^{(0)<}_{00}(t-t')+G^{({\rm ac})<}_{00}(t,t')\ ,
\end{align}
where the Fourier transform of the first term on the right-hand side is given in Eq. (\ref{G0l}), and 
\begin{align}
&G^{({\rm ac})(0)<}_{00}(\omega^{}_{\rm ac},\omega)=\frac{i\Gamma\delta\mu}{\omega^{}_{\rm ac}}
G^{(0)r}_{00}(\omega -\omega^{}_{\rm ac})G^{(0)a}_{00}(\omega )\nonumber\\
&\times [f^{}_{L}(\omega -\omega^{}_{\rm ac})-f^{}_{L}(\omega )-
f^{}_{R}(\omega -\omega^{}_{\rm ac})+f^{}_{R}(\omega )]\ .
\label{G0acl}
\end{align}
The Green's function Eq. (\ref{G0acl}) is written for a spatially-symmetric junction, $\Gamma_{L}=\Gamma_{R}=\Gamma/2$ [see Eq. (\ref{Gamma})]. The relevant diagrams (whose expressions are given in Appendix \ref{detC}) are depicted in Fig. \ref{Fig8}.

\begin{widetext}

\begin{figure}[htp]
\includegraphics[width=6cm]{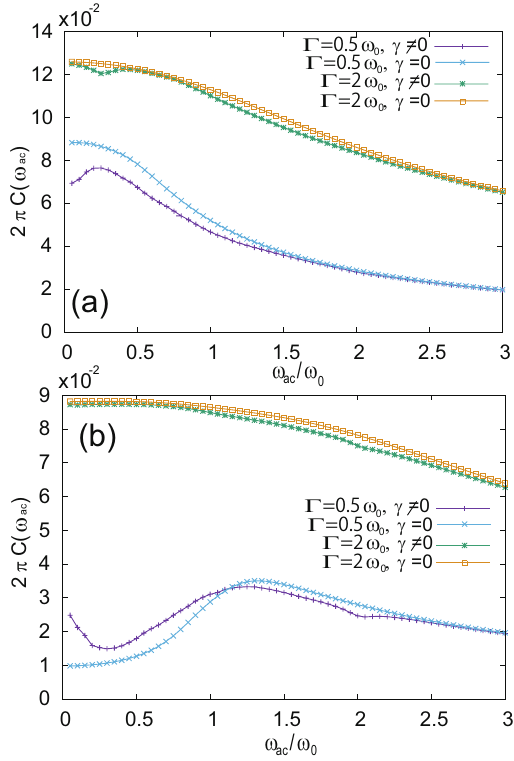}\hspace{2cm}\includegraphics[width=6.3cm]{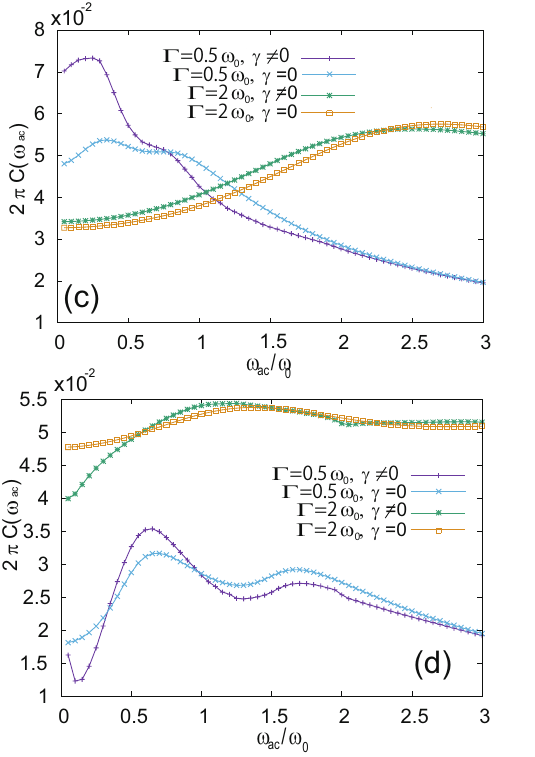}
\caption{(color online) 
The zero-temperature  transport coefficient $C(\omega_{\rm ac})$ [Eq. (\ref{Ct})]  
as  a function of the ac-field frequency, measured in units of $\omega_{0}$, for various values of the junction transmission  ${\cal T}$ and the bias voltage $V$. (a) ${\cal T}=0.8$, $eV=0.5\omega_{0}$; (b) 
${\cal T}=0.2$, $eV=0.5\omega_{0}$; (c)
${\cal T}=0.8$, $eV=2\omega_{0}$; (d) 
${\cal T}=0.2$, $eV=2\omega_{0}$. There are two pairs of curves in each panel. The members of each pair correspond to the same value of the resonance width (either $\Gamma=0.5\omega_{0}$ or $\Gamma=2\omega_{0}$); one of them is plotted for zero electron-vibration coupling, and the other for  $\gamma=0.5 \omega_{0}$.  
}
\label{Fig3}
\end{figure}
\end{widetext}

The transport coefficient $C(\omega_{\rm ac})$, in the presence and in the absence of the electron-vibration interaction,  is plotted in Fig. \ref{Fig3} as a function of $\omega_{\rm ac}/\omega_{0}$.
Panels (a) and (b) there  show it for a high-transparency junction, ${\cal T}=0.8$, 
for $eV=0.5\omega_{0}$ and $eV=2\omega_{0}$, respectively.
The other parameters are   $\mu-\epsilon_{0}=0.125 \omega^{}_{0},  \Gamma=0.5\omega_{0}$
for the upper pair of curves, and 
$\mu-\epsilon_{0}=0.5 \omega^{}_{0},  \Gamma=2\omega_{0}$ for the lower pair. 
The coupling of the charge carriers  to the vibrations suppresses 
$C(\omega_{\rm ac})$  for both pairs of parameters when the bias voltage is low, $eV=0.5\omega_{0}$.
The coupling to the vibrations has almost no effect on $C(\omega_{\rm ac})$ as the  ac frequency exceeds the vibrations' frequency; one might say that   
the vibrations cannot ``catch up" there.
For the higher bias voltage, $eV=2\omega_{0}$, 
the coupling with the vibrations causes $C(\omega_{\rm ac})$ to diminish for the large resonance width, $\Gamma=2\omega_{0}$, 
but enhances it for the smaller value of $\Gamma(=0.5\omega_{0})$ when $\omega_{\rm ac}<\omega_{0}$.
Panels (c) and (d) in Fig. \ref{Fig3} display $C(\omega_{\rm ac})$ for a low-transmission junction, 
${\cal T}=0.2$.
Here, 
 $\mu-\epsilon_{0}=0.5 \omega^{}_{0},  \Gamma=0.5\omega_{0}$
for  one pair of curves, and 
$\mu-\epsilon_{0}=2 \omega^{}_{0},  \Gamma=2\omega_{0}$ for the other. 
It is seen  that the coupling to the vibrations enhances  $C(\omega_{\rm ac})$ at small ac frequencies for $eV=0.5\omega_{0}$ [Fig. \ref{Fig3}(c)], but decreases it  for the higher ones: $0.8\omega_{0}\lesssim\omega_{\rm ac}$
for $\Gamma=0.5\omega_{0}$
and 
$2.3\omega_{0}\lesssim\omega_{\rm ac}$
for $\Gamma=2\omega_{0}$. 
At $eV=2\omega^{}_{0}$ 
[Fig. \ref{Fig3}(d)]
$C(\omega_{\rm ac})$ 
diminishes when $\omega_{\rm ac}\lesssim 0.5\omega_{0}$ for both values of $\Gamma$, 
but increases for $\omega_{\rm ac}\sim 0.7\omega_{0}$
and $\Gamma=0.5\omega_{0}$, and 
for $\omega_{\rm ac}\sim \omega_{0}$
and $\Gamma=2\omega_{0}$, respectively. Again we find that $C(\omega_{\rm ac}) $ tends to decrease gradually 
at high ac frequencies.
Appendix \ref{detC}  examines the contribution of each diagram separately to $C(\omega_{\rm ac})$, see Figs. 
\ref{Fig10_8} and \ref{Fig10_2}.

\subsection{Features of the vibrations under an ac field}
\label{vib}

Within the framework of the Hamiltonian Eq. (\ref{HAM}), the dynamics of the vibrational modes is brought about via their coupling with the charge carriers; they are affected by the ac field also only through that interaction. Technically,  the dynamics of the vibrations is described by the particle-hole propagator, see the diagrams in Fig. \ref{Fig8}. Here we focus our attention on  the ac field-induced  modifications of two quantities. The first  comes  from the correlation $
d^{<}(t,t')=-i\langle b^{\dagger}(t')b^{}(t)\rangle$, the second from the usual vibration Green's functions, given in Eq. (\ref{Dcon}).

Formally, the distribution of the vibrational modes is given by the equal-time correlation
\begin{align}
N(t)=id^{<}_{}(t,t), \ \ d^{<}(t,t)=-i\langle b^{\dagger}(t)b^{}(t)\rangle\ .
\label{defN}
\end{align}
Note that this definition does not coincide with the one given in Refs. \onlinecite{Avriller,Kaasbjerg}, which  uses the equal-time Green's function $D$ [Eq. (\ref{Dcon})], that is $N(t)=[-1+iD^{<}(t,t)]/2$.   The reason is the presence of the ac field which causes $\langle [b^{\dagger}(t)]^{2}\rangle $ and $\langle [b(t)]^{2}\rangle $ to be nonzero. 
The equal-time conventional vibration Green's function $D^{<}(t,t)$ [see Eq. (\ref{Dcon})] 
yields the fluctuation of the displacement of the harmonic oscillator representing the junction, $(\Delta X)^{2}$, as  the coordinate operator $\hat{X}$ along the junction (assumed to lie along the ${\bf x}$ direction)
is given by 
\begin{align}
\hat{X}\sqrt{2m\omega_{0}^{}}=(b+b^{\dagger})\ , 
\end{align}
where $m$ is the mass of the oscillator representing the molecule.
Hence
\begin{align}
iD^{<}_{}(t,t)=2m\omega^{}_{0}[\Delta X(t)]^{2}_{}\ .
\label{XD}
\end{align}
We expand both  quantities, Eq. (\ref{defN}) and Eq. (\ref{XD}),  to linear  order in the ac  chemical potentials, to obtain (for the Fourier transforms at the ac frequency)
\begin{align}
N(\omega^{}_{\rm ac})&=N^{(0)}_{}+N^{(1)}_{L}(\omega^{}_{\rm ac})\delta\mu^{}_{L}
+
N^{(1)}_{R}(\omega^{}_{\rm ac})\delta\mu^{}_{R}\ , 
\label{Nw}
\end{align}
and 
\begin{align}
&[\Delta X^{}_{}(\omega^{}_{\rm ac})]^{2}_{}=[\Delta X^{(0)}_{}(\omega^{}_{\rm ac})]^{2}_{}\nonumber\\
&+
[\Delta X^{(1)}_{L}(\omega^{}_{\rm ac})]^{2}_{}\delta\mu^{}_{L}+
[\Delta X^{(1)}_{R}(\omega^{}_{\rm ac})]^{2}_{}\delta\mu^{}_{R}\ .
\label{Xw}
\end{align}
The first terms on the right  hand-side of Eqs. (\ref{Nw}) and (\ref{Xw}) refer to the case where there is no ac field.

Below, we present results for the sum of the two linear-response coefficients, 
\begin{align}
N^{(1)}_{}(\omega^{}_{\rm ac})=
N^{(1)}_{L}(\omega^{}_{\rm ac})+
N^{(1)}_{R}(\omega^{}_{\rm ac})\ ,
\label{popc}
\end{align}
and 
\begin{align}
\Delta X^{(1)}_{}(\omega^{}_{\rm ac})=\Big [
[\Delta X^{(1)}_{L}(\omega^{}_{\rm ac})]^{2}+
[\Delta X^{(1)}_{R}(\omega^{}_{\rm ac})]^{2}\Big ]^{1/2}_{}\ .
\label{xd}
\end{align}
Both quantities are complex, since as mentioned the calculations use for the ac chemical potentials the form $\delta\mu_{L(R)}\exp[i\omega_{\rm ac}t]$; the phase delay of $N^{(1)}$ and $\Delta X^{(1)}$ yield the phase shift away from $\omega_{\rm ac}t$. We plot below the absolute values of $N^{(1)}_{}(\omega^{}_{\rm ac})$ and of  $\Delta X^{(1)}_{}(\omega^{}_{\rm ac})$, and the phase  of the latter (see Ref. \onlinecite{Imamura}).

The absolute value of $N^{(1)}$ is plotted in Fig. \ref{Fig4}(a)  for a transparent junction whose transmission is 0.8, and in Fig. \ref{Fig5}(a) for an opaque junction, with ${\cal T}=0.2$.
Perhaps not surprisingly, the larger is the bias voltage $V$, the larger is 
$|N^{(1)}|$, for both types of junctions. This  tendency was found also in the absence of the ac field. \cite{OEW.2010}
Large values of $\omega_{\rm ac}/\omega_{0}$ suppress the effect of the ac field on the vibrations' occupation.

\begin{figure}[htp]
\includegraphics[width=7cm]{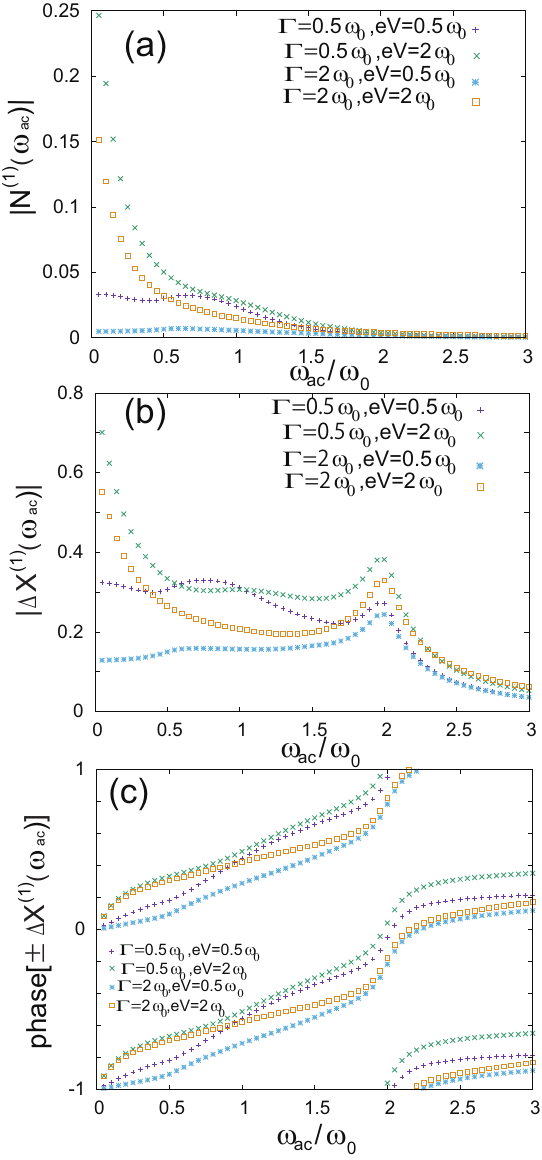}
\caption{(color online) A transparent junction, ${\cal T}=0.8$.  (a) The change in the vibrations' population due to the ac field,
[$|N^{(1)}|$,  see Eq. (\ref{popc})]       scaled by $\omega^{-1}_{0}$, as a function of $\omega_{\rm ac}/\omega_{0}$; (b)  the change in the displacement's fluctuation amplitude [in units of $(2m\omega_{0})^{-1/2}$, see Eq. (\ref{xd})],  due to the ac field as a function of $\omega_{\rm ac}/\omega_{0}$; (c) the phase delay (in units of $\pi$) of the displacement's fluctuation.}
\label{Fig4}
\end{figure}

\begin{figure}[htp]
\includegraphics[width=7cm]{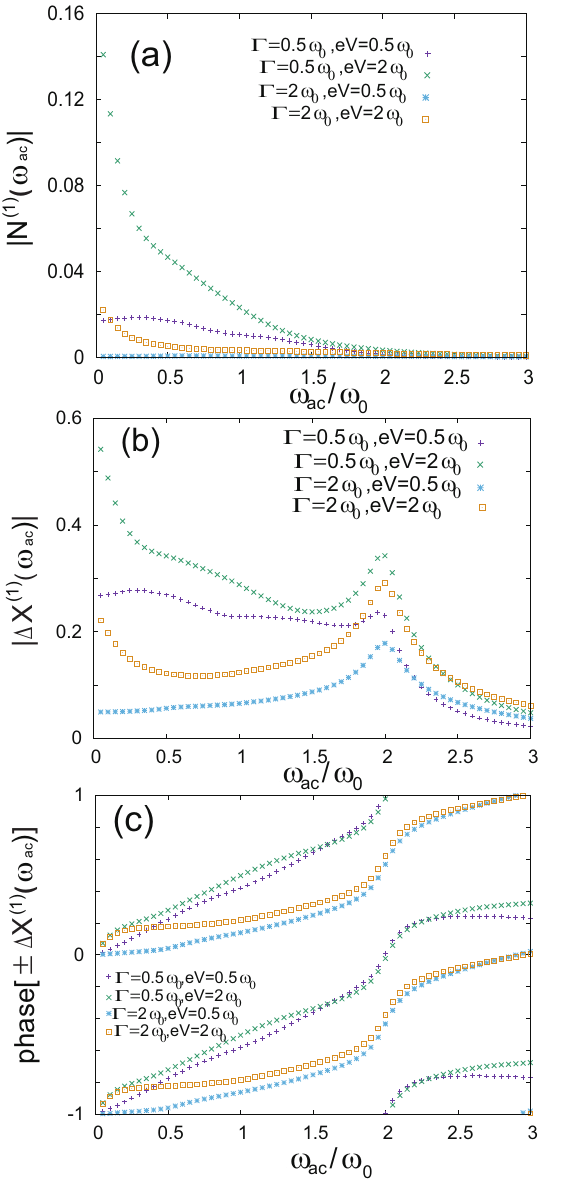}
\caption{(color online) 
An opaque junction, ${\cal T}=0.2$ (the other parameters are as in Fig. \ref{Fig4}). 
}
\label{Fig5}
\end{figure}

While the  ac field-modified amplitude $|\Delta X^{(1)}|$, Fig. \ref{Fig4}(b)  for the high transmission  and Fig. \ref{Fig5}(b) for the lower one, also decays at high values of the ac frequency, this fluctuation possesses  a distinct 
peak around $\omega_{\rm ac}\sim 2\omega_{0}$, apparently of the same origin as the dip in the contribution of the two-vibrational modes processes in the particle-hole propagator to $C(\omega_{\rm ac})$, see Figs. \ref{Fig10_8}(c) and \ref{Fig10_2}(c).
The phase delay of $\Delta X^{(1)}$, Eq. (\ref{xd}), is a rather interesting property. The displacement $x$  of a classical forced-oscillator obeys the equation of motion
\begin{align}
\ddot{x}+\eta\dot{x}+\omega^{2}_{0}x=f\cos(\omega^{}_{\rm ac}t)\ ,
\end{align}
where $f$ is the force (normalized to the proper units).
The amplitude $A$ of this driven operator is 
\begin{align}
A=\frac{f}{\sqrt{(\omega^{2}_{0}-\omega^{2}_{\rm ac})^{2}
+\eta^{2}_{}\omega^{2}_{\rm ac}}}\ ,
\end{align}
and its phase shift $\theta$ is
given by
\begin{align}
{\rm tan}\theta=\frac{\eta\omega^{}_{\rm ac}}{\omega^{2}_{0}-\omega^{2}_{\rm ac}}\ .
\end{align}
Thus at $\omega_{\rm ac}=\omega_{0}$,  the phase delay of the classical oscillator  reaches  $\pi/2$ and ``jumps" from being positive ($\omega^{2}_{0}>\omega^{2}_{\rm ac}$) to a negative value.
As seen in Figs. \ref{Fig4}(c) and \ref{Fig5}(c), 
the phase delay of the displacement  is 0 or $-\pi$ at very small ac frequencies, 
and reaches the out-of-phase configuration
where it is $\pm\pi/2$
around $\omega_{\rm ac}\gtrsim \omega_{0}$. The fact that it occurs at higher frequencies,  $\omega_{\rm ac}>\omega_{0}$, (e.g., for $\Gamma>\omega_{0}$)
indicates that the resonance itself is shifted. In any event, one may understand this dependence on classical grounds. This is not the case for the additional structure at $\omega_{\rm ac}\sim 2\omega_{0}$. As observed for the amplitude $|\Delta X^{(1)}|$, this feature is related 
to the two-vibrational modes processes contributing to the particle-hole propagator (see Appendix \ref{detC});
it cannot be explained by referring to the classical driven oscillator. This feature is thus a manifestation of an electro-mechanical interaction in the quantum regime.

 \section{\MakeLowercase{dc}-current noise and full-counting statistics}
 \label{fcs}

The term full-counting statistics (FCS) refers to the  distribution function of the probability
$P_{\tau}(q)$  for the number $q$  of transmitted charges to traverse  a quantum conductor during a certain time, $\tau$,  at out-of-equilibrium conditions. 
We consider the FCS when   no ac field is applied to the junction,  $\delta\mu_{L}=\delta\mu_{R}=0$ in Eq. (\ref{mu}), and thus
\begin{align}
\mu_{L}-\mu_{R}=eV\  , 
\label{V}
\end{align}
where $V$ is the bias voltage (the frequency-dependent third cumulant  
is considered in Ref. \onlinecite{Emary}).
In our previous work, \cite{AU.2013}
we have investigated the FCS for weak electron-vibration coupling, emphasizing in particular the fluctuation theorem (FT) and charge conservation. 
The FT has been recently applied to nonequilibrium quantum transport, \cite{Tobiska,Forster,Saito,Utsumi,Andrieux,Esposito,Campisi,Altland,Lopez,Utsumi.2010} including the present issue in the strong-coupling regime. \cite{Simine,Schaller}
The FT is a consequence of micro-reversibility and can be understood as a microscopic extension of the second law of thermodynamics.   It relates the probabilities to find negative and positive  entropy productions, 
\begin{align}
P^{}_{\tau}(-q)=P^{}_{\tau}(q)e^{-\beta qeV}\ .
\label{P}
\end{align}
Despite its simple appearance,  Eq. (\ref{P}) 
produces linear-response results, i.e., it ensures the fluctuation-dissipation theorem and Onsager's reciprocal relations close to equilibrium,  \cite{Tobiska,Forster,Saito,Utsumi,Andrieux,Esposito,Campisi}
while conveying invaluable information at nonequilibrium conditions.

The Fourier transform of the probability distribution is connected to the Keldysh partition function \cite{Keldysh}
\begin{align}
&\sum_{q}P^{}_{\tau}(q)e^{iq\lambda}=\Big \langle \tilde{T}
\exp[i\int_{0}^{\tau}dt {\cal H}^{}_{{\rm tun},-}(t)^{}_{I}]\nonumber\\
&\times
T
\exp[-i\int_{0}^{\tau}dt {\cal H}^{}_{{\rm tun},+}(t)^{}_{I}]\Big \rangle \approx \exp [\tau{\cal F}(\lambda)]\ ,
\label{C}
\end{align}
where $T$ ($\tilde{T}$) is the (anti) time-ordering  operator. 
The subscript $I$ indicates that time dependency is in the interaction picture [see Eq. (\ref{hamtun})].

 At steady state, that is, in  the long measurement-time limit, $\tau\rightarrow\infty$,  the  
 cumulants of the current, e.g., the noise,  are derived from the derivatives of the cumulant generating-function (CGF), ${\cal F}$, 
 \begin{align}
\llangle I^{n}_{}\rrangle =e^{n}\frac{\partial^{n}_{}{\cal F}(\lambda)}{\partial (i\lambda)^{n}_{}}\Big |^{}_{\lambda=0}\ .
\label{In}
 \end{align}
As mentioned, our calculations are confined to the regime of a weak  electron-vibration coupling energy, and therefore require an expansion in  $\gamma$. However, there is a subtle point in this procedure.  A na\"{i}ve  second-order perturbation theory is not capable of producing the correct nonequilibrium distribution function of the vibrations. \cite{Haupt,OEW.2010}
 One has therefore to re-sum an infinite number of diagrams by adopting the linked cluster expansion (see, e.g., Ref. \onlinecite{Utsumi.2006}), or to use an even more advanced method, the nonequilibrium Luttinger-Ward functional,  as exploited in Ref. \onlinecite{AU.2013}. 
In terms of this functional, the CGF is
\begin{align}
{\cal F}(\lambda)={\cal F}^{}_{0}(\lambda)-\overline{\Phi}^{(2)}_{}(\lambda)\ ,
\label{calF}
\end{align}
where ${\cal F}_{0}$ is the CGF of the noninteracting electrons (i.e., the CGF for $\gamma=0$), 
\begin{align}
{\cal F}^{}_{0}(\lambda)=\frac{1}{2\pi\beta}\Big [({\rm arccosh}X^{}_{\lambda})^{2}
\Big ]
\ ,
\label{F0}
\end{align}
with
\begin{align}
X^{}_{\lambda}=(1-{\cal T}){\rm cosh} \frac{\beta eV}{2}+{\cal T}{\rm cosh}\frac{\beta eV+2i\lambda}{2}\ , 
\end{align}
where  ${\cal T}$ is the transmission of the localized level. 
A trivial constant preserving the normalization condition, ${\cal F}_{0}(0)=0$, is omitted here for convenience.

The effect of the electron-vibration interaction on the FCS is contained in the second term 
of Eq. (\ref{calF}). Using a diagrammatic expansion in powers of the small parameter 
\begin{align}
g=2\gamma^{2}/(\pi\Gamma^{2})\ ,
\label{g}
\end{align}
where $\Gamma$, Eq. (\ref{Gamma}), is the width of the resonance,  
Ref. \onlinecite{AU.2013} shows
that $\overline{\Phi}^{(2)}$ is faithfully described within the    random-phase approximation (RPA),
\begin{align}
\overline{\Phi}^{\rm RPA}_{}(\lambda)=\overline{\Phi}^{(2)}_{}(\lambda)+{\cal O}(g^{2}_{})\ , 
\end{align}
up to second order in $\gamma$, i.e, up to first order in $g$, Eq. (\ref{g}).
In terms of the vibration Green's function $D_{\lambda}(\omega)$,
\begin{align}
\overline{\Phi}^{\rm RPA}_{}(\lambda)=\frac{1}{4\pi}\int d\omega \ln{\rm det} D^{-1}_{\lambda}(\omega)\ .
\end{align}
Here, 
\begin{align}
D^{-1}_{\lambda}(\omega)=\left [\begin{array}{cc}
\frac{\omega^{2}_{}-\omega^{2}_{0}}{2\omega^{}_{0}}-\Pi^{++}_{\lambda}(\omega)&\Pi^{+-}_{\lambda}(\omega )\\
\Pi^{-+}_{\lambda}(\omega)&
\frac{\omega^{2}_{0}-\omega^{2}_{}}{2\omega^{}_{0}}-\Pi^{--}_{\lambda}(\omega)\end{array}\right ]\ ,
\label{D}
\end{align}
 where
$\Pi$  is  the
 electron-hole propagator, resulting from the  polarization of the electrons [Eqs. (\ref{pra}) and (\ref{pl}), modified in a non-trivial way to include the effect of the counting field $\lambda$, see Ref. \onlinecite{AU.2013} for details].

\begin{figure}[htp]
\includegraphics[width=7cm]{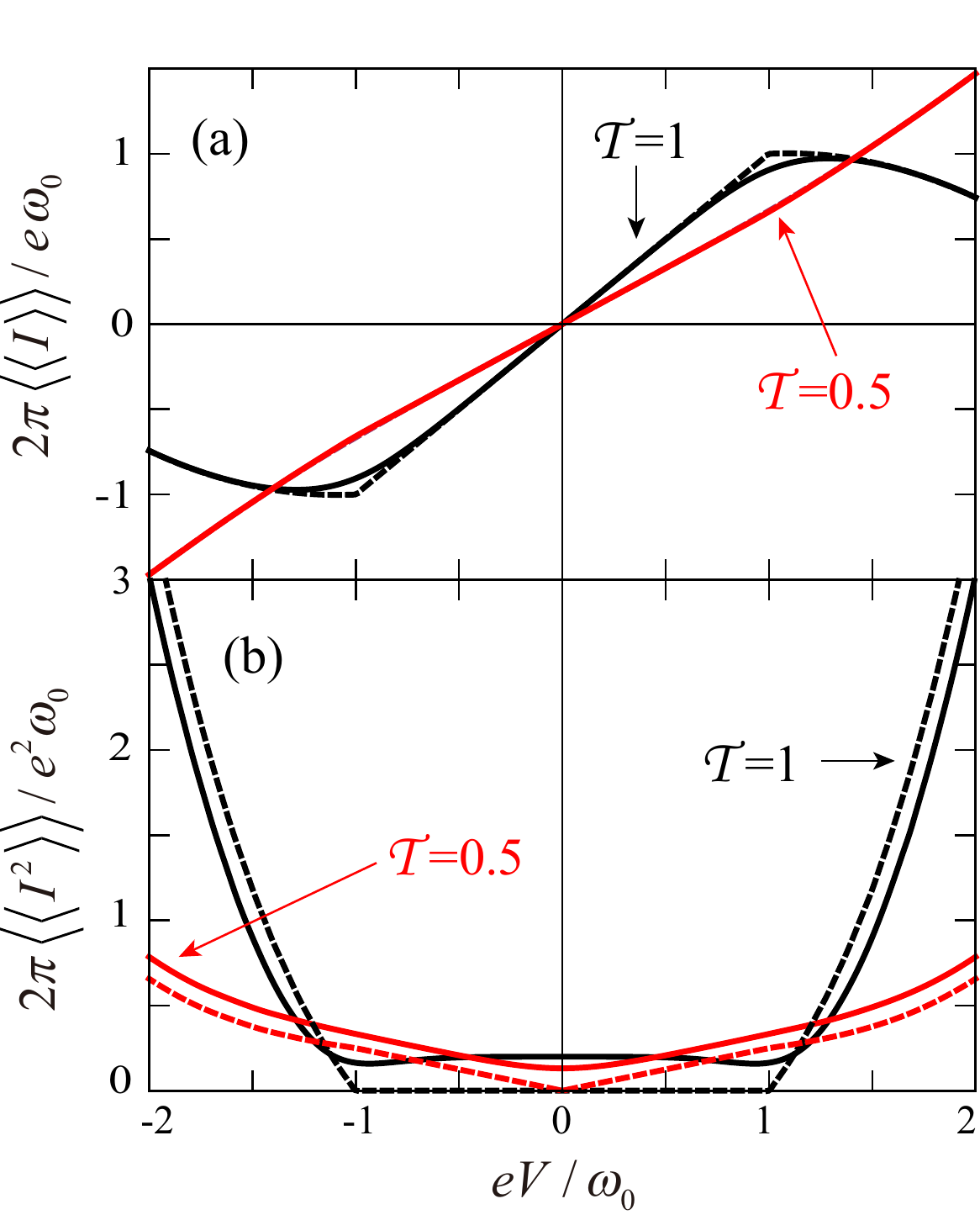}
\caption{(color online) The source-drain bias voltage dependence of the current  (a), and  of the current noise (b), for a perfect transmission,  ${\cal T}=1$ (black curves), and for ${\cal T}=0.5$ (red curves). The solid curves are for $\beta\omega_{0}=10$, and the dashed ones are for zero temperature. The electron-vibration coupling constant $g$ [Eq. (\ref{g})] is 0.1. Adapted with permission from Ref. \onlinecite{AU.2013}; copyrighted by the American Physical Society. }
\label{Fig6}
\end{figure}

Figure \ref{Fig6}(a) displays  the average current  at a finite temperature (solid curves)  and at zero temperature (dashed curves), as a function of the bias voltage, for a spatially-symmetric junction.  (Temperature is measured in units of $\omega_{0}$ and the bias is scaled by $\omega_{0}/e$.)   When  the transmission is perfect,  ${\cal T}=1$, the current is suppressed  
once 
$|eV|>\omega_{0}$, since the electrons can then be backscattered inelastically by the vibrations.
At weaker transmissions, e.g., ${\cal T}=0.5$, 
 the current is slightly enhanced above this threshold. \cite{Haupt} Apparently in this regime  the main effect of the  inelastic scattering of the electrons by the vibrations is to add more  transport channels and to  increase  the phase-space volume available for the scattering events. 
A finite temperature tends to smear the kink structure of the ${\cal T}=1$ curve. The effect of the temperature on the current for ${\cal T}=0.5$   
 is rather insignificant.

Figure \ref{Fig6}(b) depicts
 current noise, $\llangle I^{2}\rrangle$, Eq. (\ref{In}) with $n=2$. At perfect transmission and zero temperature,   the  noise  vanishes below the threshold, at $|eV|<\omega_{0}$. Thermal fluctuations which arise at a finite temperature induce additional noise there. Although the current, i.e., $\llangle I\rrangle$,  is suppressed above the threshold,  the  inelastic  scattering resulting from the interaction of the charge carriers  with the vibrations   
 enhances the noise in that regime.   As in Fig. \ref{Fig6}(a), the effect of the temperature on the noise in an imperfect junction (${\cal T}=0.5$), is far less dramatic, the noise is simply enhanced.

\begin{figure}[htp]
\includegraphics[width=7cm]{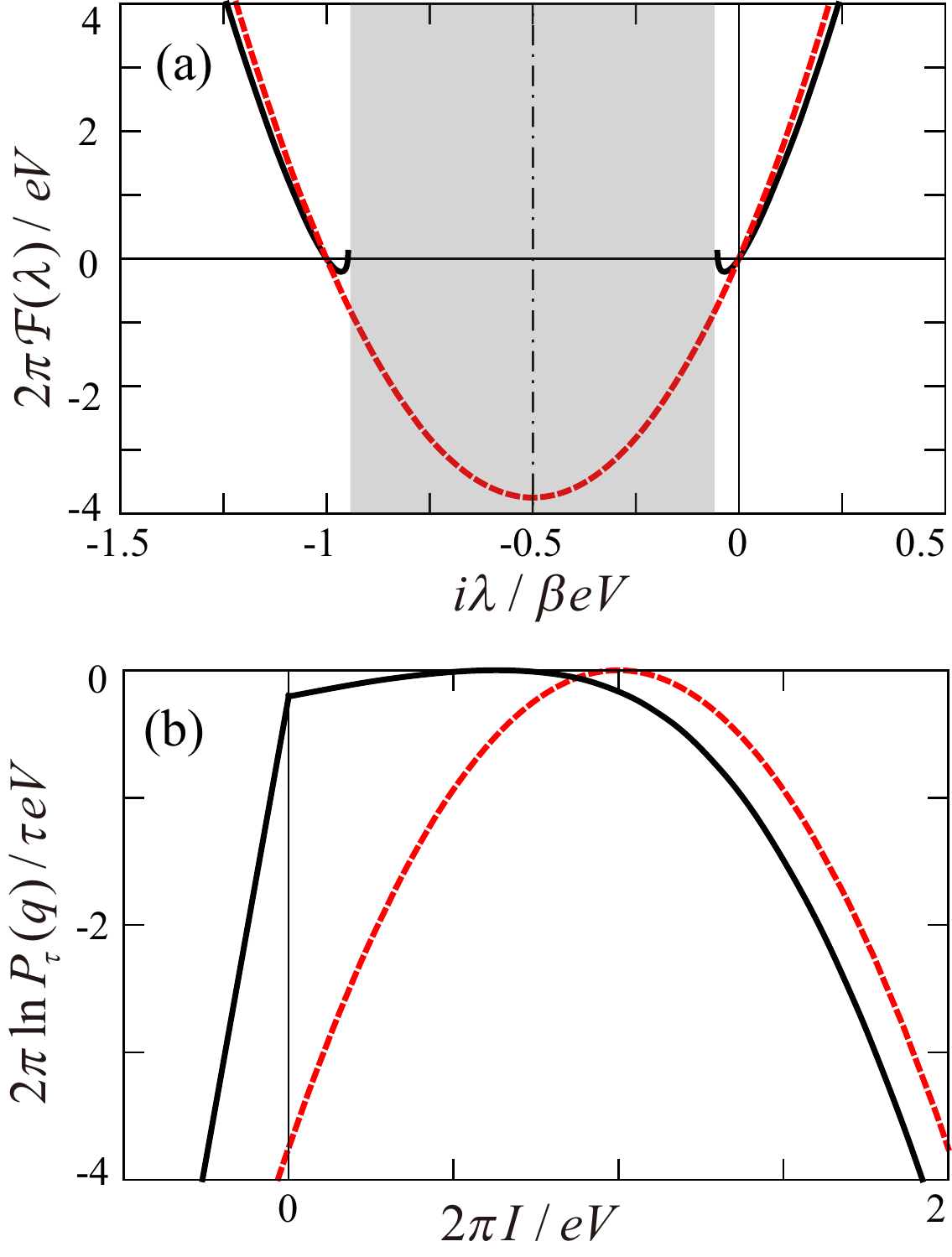}
\caption{The CGF  [as a function of $i\lambda/(\beta eV)$] (a), and the rate function [as a function of $I\propto q/\tau/(eV/(2\pi))$] (b),  for a perfect transmission, ${\cal T}=1$, $eV/\omega^{}_{0}=1.5$ and $\beta\omega^{}_{0}=10$. The solid (black) and the  dotted (red) curves show results with (g=0.1) and without (g=0) the electron-vibration coupling, respectively. In the dashed area of panel (a) the CGF of interacting electrons is non-analytic and non-convex, resulting in a non-differentiable point of the rate function at $I=0$ (b). Figure adapted with permission from Ref. \onlinecite{AU.2013}; copyrighted by the American Physical Society.}
\label{Fig7}
\end{figure}

As mentioned, the CGF is constructed from the long-time limit of the probability distribution $P_{\tau}(q)$ of 
a charge $eq$ passing through the junction during a time $\tau$. 
The approach of this probability to its steady-state value is characterized by the rate function, defined as $-\lim_{\tau\rightarrow\infty} \ln P_{\tau}(\tau I)/\tau$, where in the long-time limit the charge $eq$  is scaled as $I\tau$. 
 Figure \ref{Fig7} displays the CGF and the rate function at perfect transmission. For comparison, we plot the corresponding curves for noninteracting electrons. \cite{AU.2013}  The width of the rate function is enhanced by inelastic phonon scattering [Fig. \ref{Fig7}(b)]. The CGF obeys the fluctuation theorem:  the curves are symmetric around the dot-dashed vertical line at $i\lambda =-\beta eV/2$ [Fig. \ref{Fig7}(a)]. The peak of the probability distribution is shifted in the negative direction and the probability to find large current fluctuations is suppressed as compared with that pertaining to noninteracting electrons  [Fig. \ref{Fig7}(b)]. In the shaded area of Fig. \ref{Fig7}(a), the CGF is non-analytic and non-convex. Correspondingly, the rate function has a non-differentiable point at $I=0$ [see Fig. \ref{Fig7}(b)]. As a result, although the probability to observe currents smaller than the average value is enhanced by the inelastic  scattering due to the coupling with the vibrations, the probability 
 to find negative currents $I<0$ is strongly reduced. This is consistent with the fluctuation theorem, which states that although thermal agitations generate current flowing in the opposite direction to the source-drain bias, that probability is exponentially suppressed at low temperatures. 
 
The analogy between the characteristic function Eq. (\ref{C}) and the equilibrium partition function helps to utilize 
statistical-physics methods for characterizing statistical properties of current fluctuations. The large-deviation approach connects the non-convexity  in Fig. \ref{Fig7}(a) and the kink in Fig. \ref{Fig7}(b) with those observed in the Helmholtz free energy and the Gibbs free energy  at the liquid-gas first-order phase transition. For noninteracting particles, zeros of the characteristic function (\ref{C}) reside on the negative real axis in the complex $z=\exp[i\lambda]$ plane,  \cite{Abanov} reminiscent of the Yang-Lee zeros. \cite{Yang} The zeros'  distribution 
can be used to characterize the interaction.  \cite{Flindt}
In Ref. \onlinecite{AU.2013}
it is found that the distribution of the singularities of ${\cal F}$ in the complex $\lambda-$plane can be a useful tool for identifying the effect of the electron-vibration interaction on the current probability distribution.

\section{Conclusions}
\label{conclu}

Our work is centered on  nonequilibrium features in the dynamics of molecular junctions, in the regime of weak electron-vibration coupling.  Specifically, we have presented a detailed study of the response of the junction to a weak ac field, in the presence of an arbitrary bias voltage (constant in time).  In Sec. \ref{basic} we have analyzed the electronic response of the charge carriers to an ac field, and the modifications introduced by that field in the vibrations' distribution and in the fluctuations of the 
coordinate of the center of mass of the junction, as a function of the ac frequency $\omega_{\rm ac}$. Perhaps the most intriguing feature we find is the (relatively) sharp structure around $\omega_{\rm ac}\sim 2\omega_{0}$ ($\omega_{0}$ is the frequency of the vibrational mode). While the structure around $\omega_{\rm ac}\sim \omega_{0}$ can be understood on a classical ground, the one at $\omega_{\rm ac}\sim 2\omega_{0}$ is related to quantum effects in the particle-hole propagator (see Secs. \ref{ac} and \ref{vib}, and Appendix \ref{detC}). It is thus a manifestation of an electro-mechanical feature in coherent transport through a single-channel molecular junction. The effect of the arbitrary bias voltage is dwelled upon in particular in Secs. \ref{dc} and \ref{fcs}. Here we identify a change of behavior of the dc current and its noise   when the voltage is around $\omega_{0}$ (using units in which $\hbar=1$ and $e=1$). We hope that a future study of the full-counting statistics and the cumulant generating-function when the junction is also placed in an ac field will substantiate our understanding of the dynamics of the excitations in molecular junctions.


\begin{acknowledgments}
We  thank Hiroshi Imamura for useful discussions. This work was  supported by JSPS KAKENHI Grants No. JP26220711 and No. 6400390,  by  the Israel Science Foundation
(ISF),  and by the infrastructure program of Israel
Ministry of Science and Technology under contract
3-11173. 
 \end{acknowledgments}

 \appendix

\begin{widetext}

 \section{The  \MakeLowercase{ac} linear-response coefficient}
 \label{detC}

The diagrams that give the  coefficient $C(\omega_{\rm ac})$  [Eq. (\ref{Ct})],   
are depicted in Fig. \ref{Fig8}.
Figure \ref{Fig8}(a) illustrates the diagrams that contribute to $C(\omega_{\rm ac})$: 
(i) is the contribution when  the electron-vibration interaction is omitted, (ii) is that of the exchange term, and (iii) and (iv)
pertain to the vertex corrections (due to the exchange term and to the particle-hole propagator, respectively). 

In the presence of the ac field, the vibration Green's functions take the forms
\begin{align}
D^{r/a}_{}(\omega_{\rm ac}, \omega) & = D^{r/a}_{}(\omega) + D^{({\rm ac})r/a}_{}(\omega^{}_{\rm ac},\omega)\ ,
\label{a1}
\end{align}
and
\begin{align}
D^{<}_{}(\omega_{\rm ac}, \omega)  &= D^{<}_{}(\omega) + D^{({\rm ac})<}_{}(\omega^{}_{\rm ac},\omega)\ .
\label{a2}
\end{align}
The first terms on the right hand-side of Eqs. (\ref{a1}) and (\ref{a2}) are depicted in Fig. \ref{Fig8}(b), 
\begin{align}
D^{r/a}_{}(\omega)  =  D^{(0)r/a}_{}(\omega) + D^{(0)r/a}_{}(\omega) \Pi^{r/a}_{}(\omega) D^{(0)r/a}_{}(\omega)\ ,
\label{a3}
\end{align}
and
\begin{align}
D^{<}_{}(\omega) = D^{(0)<}_{}(\omega) + D^{(0)r}_{}(\omega) \Pi^{r}_{}(\omega) D^{(0)<}_{}(\omega) + D^{(0)r}_{}(\omega) \Pi^{<}_{}(\omega) D^{(0)a}_{}(\omega) + D^{(0)<}_{}(\omega) \Pi^{a}_{}(\omega) D^{(0)a}_{}(\omega)\ . 
\label{a4}
\end{align}
The second terms on the right hand-side of Eqs. (\ref{a1}) and (\ref{a2}), which are due to the ac field, are shown in Fig. \ref{Fig8}(c), 
\begin{align}
D^{({\rm ac})r/a}_{}(\omega^{}_{\rm ac},\omega)  =   D^{(0)r/a}_{}(\omega - \omega^{}_{\rm ac}) \Pi^{(ac)r/a}_{}(\omega^{}_{\rm ac}, \omega) D^{(0)r/a}_{}(\omega)\ ,
\label{a5}
\end{align}
and 
\begin{align}
D^{({\rm ac})<}_{}(\omega^{}_{\rm ac},\omega) & = 
D^{(0)r}_{}(\omega - \omega^{}_{\rm ac}) \Pi^{({\rm ac})r}_{}(\omega^{}_{\rm ac}, \omega) D^{(0)<}_{}(\omega)
+D^{(0)r}_{}(\omega - \omega^{}_{\rm ac}) \Pi_{}^{({\rm ac})<}(\omega^{}_{\rm ac}, \omega) D^{(0)a}_{}(\omega)\nonumber\\
&
+D^{(0)<}_{}(\omega - \omega^{}_{\rm ac}) \Pi_{}^{({\rm ac})a}(\omega^{}_{\rm ac}, \omega) D_{}^{(0)a}(\omega)\ .
\label{a6}
\end{align}
Here, $D^{(0)}$ is the bare vibrational modes' Green's function, 
\begin{align}
D^{(0)r/a}_{}(\omega)=\frac{1}{\omega-\omega^{}_{0}\pm i\eta/2}
-\frac{1}{\omega+\omega^{}_{0}\pm i\eta/2}\ , 
\end{align}
and
\begin{align}
D^{(0)<}_{}(\omega)=-i\Big (\frac{\eta}{(\omega+\omega^{}_{0})^{2}+(\eta/2)^{2}}(N+1)+
\frac{\eta}{(\omega-\omega^{}_{0})^{2}+(\eta/2)^{2}}N\Big )\ ,
\label{D0l}
\end{align}
where $N$ denotes the vibrations' population. Since Eq. (\ref{D0l}) is zeroth-order in the electron-vibration coupling, then in principle,  $N$ at a finite temperature is determined by the coupling to a surrounding phonon bath. \cite{OEW.2010} Here we confine ourselves to zero temperature, and hence the zeroth-order vibration Green's function contains $N=0$.  
The  
processes in which a   vibrational mode excites a particle-hole pair and then  becomes the vibrational mode again are given in Eqs. (\ref{a5}) and (\ref{a6}) [see also Eqs. (\ref{a1}) and (\ref{a2})]. The electron-hole propagator  determines the life time of the vibrational mode, when the system is effectively decoupled from an external phonon bath [see the discussion in the paragraph  following Eq. (\ref{pl})].
The third terms in these two equations refer to the  excited particle-hole pair by  the vibrational mode, under the effect of the ac field. 
The expressions for the particle-hole propagator in the absence of the ac field are given in Eqs. (\ref{pra}) and (\ref{pl}), and the ones in the presence of the ac field are 
\begin{align}
\Pi^{({\rm ac})r/a}_{}(\omega^{}_{\rm ac}, \omega)&=-i\gamma^{2}\int\frac{d\omega '}{2\pi}\Big (G^{({\rm ac})(0)<}_{00}(\omega^{}_{\rm ac},\omega ')G^{(0)a}_{00}(\omega '-\omega)+G^{(0)r}_{00}(\omega '-\omega^{}_{\rm ac})G^{({\rm ac})(0)<}_{00}(\omega^{}_{\rm ac},\omega '-\omega)\Big )\ ,
\label{pacra}
\end{align}
and
\begin{align}
\Pi^{({\rm ac})<}_{}(\omega^{}_{\rm ac}, \omega)&=-i\gamma^{2}\int\frac{d\omega '}{2\pi}\Big (G^{({\rm ac})(0)<}_{00}(\omega^{}_{\rm ac},\omega ')G^{(0)>}_{00}(\omega '-\omega)+G^{(0)<}_{00}(\omega '-\omega^{}_{\rm ac})G^{({\rm ac})(0)<}_{00}(\omega^{}_{\rm ac},\omega '-\omega)\Big )\  , 
\label{pacl}
\end{align}
where $G^{({\rm ac})(0)<}_{00}$ is given in Eq. (\ref{G0acl}) (for a symmetric junction).

\begin{figure}[htp]
\includegraphics[width=8cm]{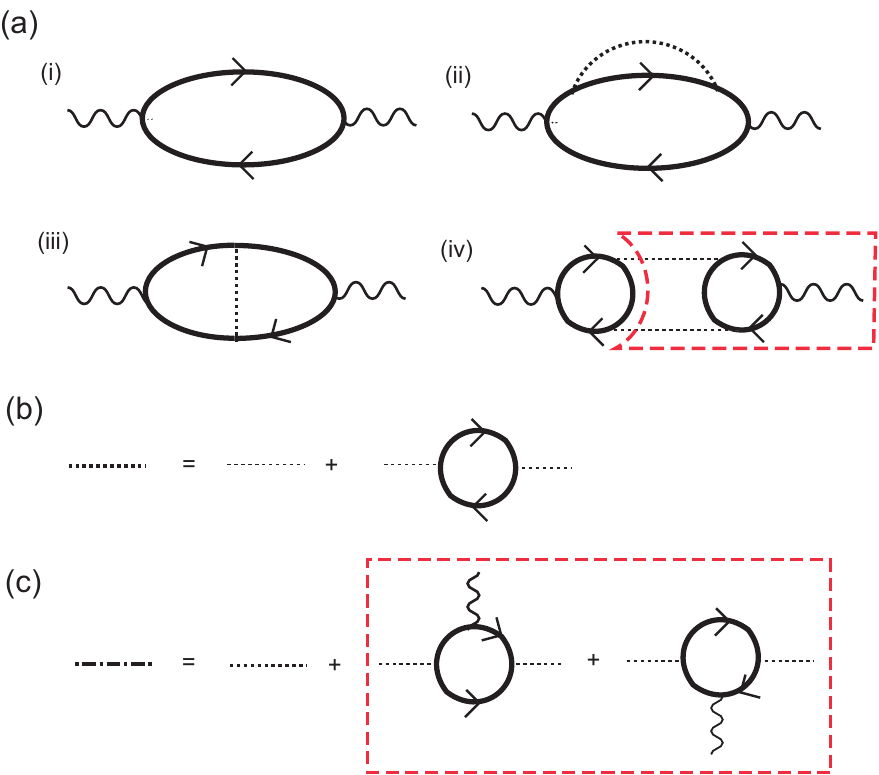}
\caption{(color online) 
(a)  The diagrams for the ac transport coefficient $C(\omega_{\rm ac})$, Eq. (\ref{Ct}). The solid lines pertain to the electron     Green's function and the dotted lines are the vibration Green's function; the wavy curve represents the ac external field.  (i) Without the electron-vibration coupling; (ii) the exchange term (without the effect of the ac field in the vibration Green's function);  (iii) vertex corrections for the exchange term  (without the effect of the ac field in the vibration Green's function); (iv)   particle-hole propagator (with the effect of the ac field in the vibration Green's function, enclosed in the dotted (red) lines).  
  (b) and (c)  The  vibration Green's function; (b) in the absence of the ac field and (c) with the effect of the ac field.  The latter, enclosed  by the dashed (red) lines,  appears due to the presence of the ac field and affect diagram (iv) in part (a).  }
\label{Fig8}
\end{figure}

For concreteness, we give the detailed expressions of the diagrams  for a spatially symmetric junction, i.e., $\Gamma_{L}=\Gamma_{R}=\Gamma/2$    in Eq. (\ref{Gamma}). Diagram (i) in Fig. \ref{Fig8}(a) pertains to the case where the electron-vibration coupling is absent, and reads
\begin{align}
C^{(0)}_{}(\omega^{}_{\rm ac})&=\frac{e}{4\omega^{}_{\rm ac}}{\rm Re}\Big [
\int\frac{d\omega}{2\pi}i\Gamma[f^{}_{L}(\omega-\omega^{}_{\rm ac})+
f^{}_{R}(\omega-\omega^{}_{\rm ac})
-f^{}_{L}(\omega)-f^{}_{R}(\omega)][G^{(0)r}_{00}(\omega-\omega^{}_{\rm ac})-G^{(0)a}_{00}(\omega )]\Big ]
\ ,
\end{align}
where the electron Green's function $G^{(0)}_{00}$ is given by Eq.  (\ref{G0ra}). The exchange contribution, diagram (ii) in Fig. \ref{Fig8}(a) reads
\begin{align}
C_{}^{({\rm ex})}(\omega^{}_{\rm ac})&=\frac{e}{4\omega^{}_{\rm ac}}
{\rm Re}\Big [
\int\frac{d\omega}{2\pi}i\Gamma 
[f^{}_{L}(\omega-\omega^{}_{\rm ac})+
f^{}_{R}(\omega-\omega^{}_{\rm ac})
-f^{}_{L}(\omega)-f^{}_{R}(\omega)][G^{({\rm ex})r}_{00}(\omega-\omega^{}_{\rm ac})-G^{({\rm ex})a}_{00}(\omega )]\Big ]
\ ,
\label{Cex}
\end{align}
where
\begin{align}
G^{({\rm ex})r/a}_{00}(\omega)=i\gamma^{2}[G^{(0)r/a}_{00}(\omega)]^{2}_{}
\int\frac{d\omega'}{2\pi}[G^{(0)<}_{00}(\omega -\omega ')D^{r/a}_{}(\omega ')+
G^{(0)r/a}_{00}(\omega -\omega ')D^{<}_{}(\omega ')
\pm
G^{(0)r/a}_{00}(\omega -\omega ')D^{r/a}_{}(\omega ')]\ .
\label{G00ex}
\end{align}
[See  Eqs. (\ref{a1}-\ref{a6}).]
The expressions for the third diagram [denoted (iii) in Fig. \ref{Fig8}(a)] and the fourth one 
[denoted (iv) in Fig. \ref{Fig8}(a)] represent vertex corrections to the transport coefficient. They both are given by 
\begin{align}
C^{({\rm ver})}_{\ell}(\omega^{}_{\rm ac})&=-\frac{e}{4\delta\mu}{\rm Re}\Big [
\int\frac{d\omega}{2\pi}i\Gamma\Big (G^{(0)r}_{00}(\omega-\omega^{}_{\rm ac})\Sigma^{({\rm ver})r}_{\ell}(\omega^{}_{\rm ac},\omega)G^{(0)r}_{00}(\omega)[f^{}_{L}(\omega)-f^{}_{R}(\omega)]\nonumber\\
&-
G^{(0)a}_{00}(\omega-\omega^{}_{\rm ac})\Sigma^{({\rm ver})a}_{\ell}(\omega^{}_{\rm ac},\omega)G^{(0)a}_{00}(\omega)[f^{}_{L}(\omega-\omega^{}_{\rm ac})-f^{}_{R}(\omega-\omega^{}_{\rm ac})]\Big )\Big ]\ .
\label{verc}
\end{align}
For the diagram (iii) in Fig. \ref{Fig8}(a) $\Sigma^{({\rm ver})r/a}_{\ell}(\omega^{}_{\rm ac},\omega)=\Sigma^{({\rm ver})r/a}_{1}(\omega^{}_{\rm ac},\omega)$, 
where
\begin{align}
\Sigma^{({\rm ver})r/a}_{1}(\omega^{}_{\rm ac},\omega)=i\gamma^{2}\int\frac{d\omega '}{2\pi}
G^{({\rm ac})(0)<}_{00}
(\omega^{}_{\rm ac},\omega -\omega ')
D^{r/a}_{}(\omega ')\ ,  
\label{sig1}
\end{align}
and consequently the contribution of that diagram is denoted $C_{1}^{({\rm ver})}$. [The
electron Green's functions appearing in Eq.  (\ref{sig1}) are given in Eq. (\ref{G0acl}).]  Diagram (iv) in Fig. \ref{Fig8}(a) is given by Eq. (\ref{verc}), with  $\Sigma^{({\rm ver})r/a}_{\ell}(\omega^{}_{\rm ac},\omega)=\Sigma^{({\rm ver})r/a}_{2}(\omega^{}_{\rm ac},\omega)$, and is denoted $C^{({\rm ver})}_{2}$,  
where
\begin{align}
\Sigma^{({\rm ver})r/a}_{2}(\omega^{}_{\rm ac}, \omega)& = i \gamma^2
\int \frac{d \omega'}{2 \pi } \Big ( [G^{(0)<}_{00}(\omega - \omega')
\pm G^{(0)r/a}_{00}(\omega - \omega')] D^{({\rm ac})r/a}_{}(\omega^{}_{\rm ac}, \omega')
+ G^{(0)r/a}_{00}(\omega - \omega') D^{({\rm ac})<}_{}(\omega^{}_{\rm ac}, \omega') \Big )
\ .
\label{sig2}
\end{align}
In the numerical  calculations the rate $\eta$ [see Eq. (\ref{dra}) and the discussion following Eq. (\ref{pl})] was chosen to be $0.1\omega_{0}$; as a result, it became larger than Im$\Pi^{a}(\omega)$ and therefore  the vibration Green's functions in Eqs. (\ref{G00ex}) and (\ref{sig1}) could be  replaced by $D^{(0)}$.


\end{widetext}

It is illuminating to study the effect of the electron-vibration interaction on each diagram separately.
Figures  \ref{Fig10_8}(a)  and  \ref{Fig10_2}(a)  display the contribution of the exchange diagram, (ii) in Fig. \ref{Fig8}(a).
This diagram represents the dressing of the electron Green's function by the  emission and absorption of vibrational modes.  At zero temperature, and when the bias voltage is smaller than $\omega_{0}$, the electrons can only absorb and emit virtual 
vibrational modes; this gives rise  to a shift in the energy of the localized level.
[This shift differs  from the  one caused by the Hartree self-energy,  Eq. (\ref{har}), which is ignored here.] The change in the localized-level energy  can increase or decrease the contribution of this diagram at small ac frequencies (as compared to $\omega_{0}$). By comparing 
Fig. \ref{Fig10_8}(a) (for  ${\cal T}=0.8$) with Fig. \ref{Fig10_2}(a) (for ${\cal T}=0.2$),  it is seen that   at $eV=0.5\omega_{0}$ and with  a small resonance width, $\Gamma=0.5\omega_{0}$, $C^{({\rm ex})}$ is negative for ${\cal T}=0.8$ and is positive for ${\cal T}=0.2$.
The opposite, but rather weaker,  tendency  is found for a wide resonance,  $\Gamma=2\omega_{0}$.
At larger biases, e.g., $eV=2\omega_{0}$, real inelastic electron-vibration scattering processes are possible. These broaden the line width of the localized level. However, modifying the resonance width and/or  the transmission  does not lead to significant effects. This is also the situation  at higher values of $\omega_{\rm ac}$. 

\begin{figure}[htp]
\includegraphics[width=7cm]{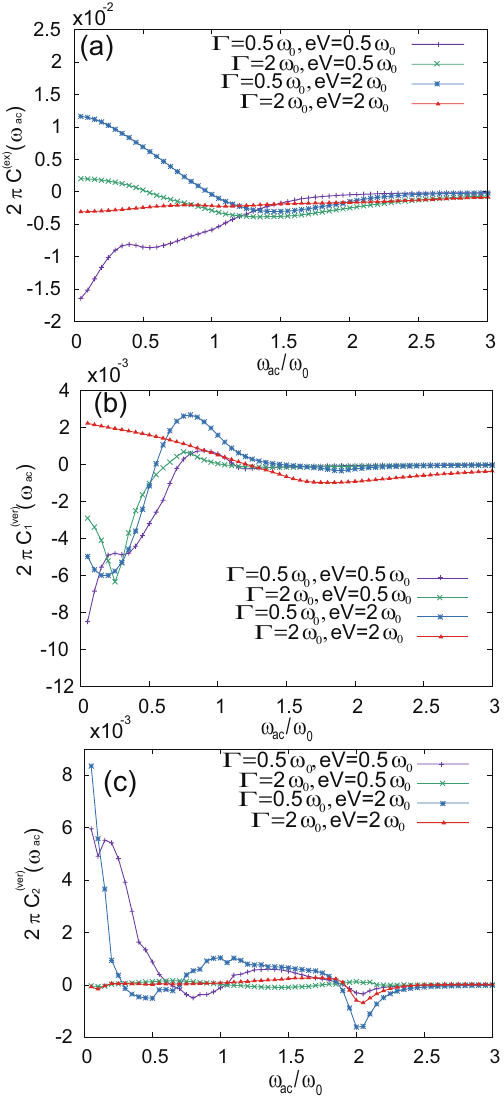}\hspace{2cm}
\caption{(color online) 
The separate contributions of the various diagrams to $C(\omega_{\rm ac})$, 
as functions of $\omega_{\rm ac}/\omega_{0}$, for   a transparent junction, ${\cal T}=0.8$. (a) $C^{({\rm ex})}$, Eq. (\ref{Cex}); (b) diagram (iii), Eq. (\ref{verc}) for $i=1$; (c) diagram (iv), Eq. (\ref{verc}) for $i=2$.}
\label{Fig10_8}
\end{figure}

The contribution of diagram (iii) in Fig. \ref{Fig8}(a) is shown in Figs. \ref{Fig10_8}(b) and  \ref{Fig10_2}(b).
This diagram corresponds to the processes in which an electron and a hole
exchange vibrational modes. This process results in a negative contribution of diagram (iii)  at small values of $\omega_{\rm ac}$, except when $\Gamma=eV=2\omega_{0}$ and the junction is transparent, Fig. \ref{Fig10_8}(b), and when
$\Gamma=eV=0.5\omega_{0}$ and the junction is opaque, Fig. \ref{Fig10_2}(b).
For both types of junctions there appears a peak in the contribution of this diagram for the larger bias voltages, $eV=2\omega_{0}$, just below $\omega_{\rm ac}=\omega_{0}$; it may reflect a resonance in the particle-hole channel  around this frequency.


\begin{figure}[htp]
\includegraphics[width=7cm]{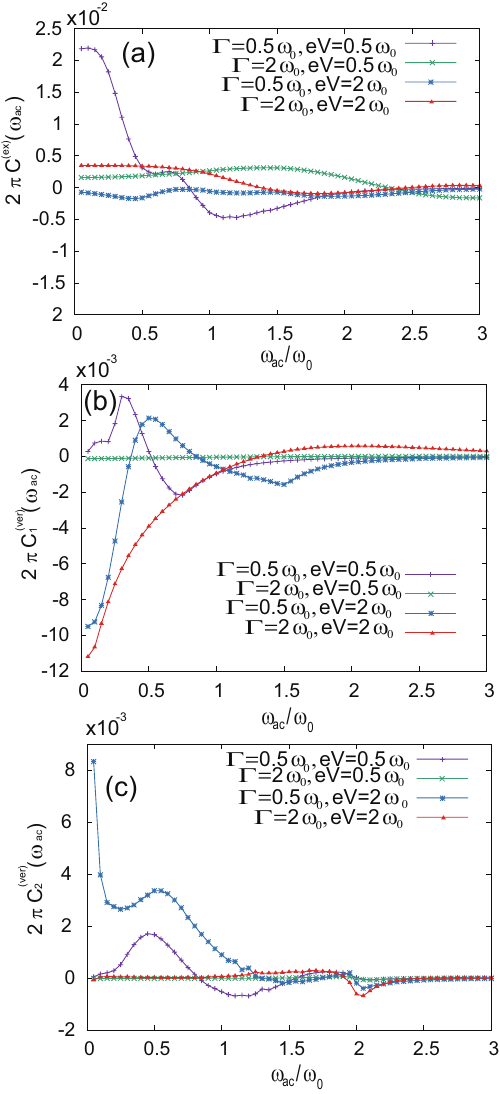}
\caption{(color online) 
Same as Fig. \ref{Fig10_8} with ${\cal T}=0.2$. 
}
\label{Fig10_2}
\end{figure}

The behavior of the contribution of diagram (iv) of Fig. \ref{Fig8}(a) is depicted in Figs. \ref{Fig10_8}(c) and \ref{Fig10_2}(c). This diagram describes two-vibration exchange between the electrons [see Fig. \ref{Fig8}(a)]. There are two distinct features in the contribution of this diagram.  First, its contribution at small ac frequencies is positive, provided that the resonance width $\Gamma$ is less than $\omega_{0}$ (otherwise, it is negligibly small). Second, it shows a dip 
 at $\omega_{\rm ac}\sim2\omega_{0}$. Apparently, the appearance of this dip, that disappears at zero bias, \cite{Imamura} can be attributed to two-phonon processes that reach resonance conditions at this frequency.

\end{document}